\documentstyle{article}

\input{tcilatex}
\newcommand{\bm}[1]{\mbox{\boldmath $#1$}}
\begin{document}

\title{A New Wave Equation For a Continuous Nondemolition Measurement}
\author{V. P. Belavkin\thanks{%
Permanent address: M.I.E.M., B. Vusovsky 3/12, Moscow 109028, USSR.} \\
Institute of Physics, Copernicus University, Toru\'{n} (Poland)}
\date{December 1988 \\
Published in: Physics Letters A, {\bf 140} (1989), No 7,8, pp 355--358}
\maketitle

\begin{abstract}
A stochastic model for nondemolition continuous measurement in a quantum
system is given. It is shown that the posterior dynamics, including a
continuous collapse of the wave function, is described by a nonlinear
stochastic wave equation. For a particle in an electromagnetic field it
reduces the Schr\"{o}dinger equation with extra imaginary stochastic
potentials.
\end{abstract}

According to the statistical interpretation of the wave function $\psi (t,%
{\bf x})$, a quantum measurement of a simple observable $\hat{X}$ of a
particle reduces it to an eigenfunction of $\hat{X}$ by a random jump
(collapse, or reduction of the wave packet). Such a jump cannot be described
by the Schr\"{o}dinger equation because the latter corresponds to the
unobserved particle. To ignore this fact gives rise to the various quantum
paradoxes of the Zeno kind. The aim of this paper is to derive a dissipative
wave equation with an extra stochastic nonlinear term describing the quantum
particle under continuous nondemolition observation. This derivation can be
obtained in the framework of quantum stochastic theory of continuous
measurements developed in refs. \cite{bib:XP31,bib:bel1,bib:ref11}, and the
quantum nonlinear filtering method recently announced in \cite{bib:ref14}.
The derivation can be done in general terms but for simplicity here we
consider a quantum spinless particle in an electromagnetic field. The
unperturbed dynamics of such a particle with mass $m$ is given by the Schr%
\"{o}dinger equation 
\begin{equation}
{\rm i}\hbar \partial \psi /\partial t=[({\bf U}+{\rm i}\hbar {\bf \nabla }%
)^{2}/2m+V]\psi ,  \label{eq:1}
\end{equation}%
where ${\bf U}=e{\bf A}/c$ and $V=e\Phi $, ${\bf A}(t,{\bf x})$ and $\Phi (t,%
{\bf x})$ are the scalar and vector field potentials.

A continuous process $\dot{{\bf Y}}$ is taken as the sum 
\begin{equation}
\dot{{\bf Y}}(t)=(2\lambda )^{1/2}\hat{{\bf X}}(t)+{\bf e}(t)  \label{eq:2}
\end{equation}%
of the standard white noise ${\bf e}=(e_{1},e_{2},e_{3})$ (error), 
\[
\langle e_{i}(r)e_{k}(s)\rangle =\delta _{ik}\delta (r-s),
\]%
and the particle coordinate operator process ${\bf X}=(X_{1},X_{2},X_{3})$
in the Heisenberg picture $\hat{X}(t)=U(t)^{\ast }XU(t)$ with amplification $%
\sqrt{2\lambda }$ ($\lambda >0$ is the accuracy coefficient).

Indirect measurement of the position of the particle described by (\ref{eq:2}%
) perturbs its dynamics (\ref{eq:1}) in such a way that the vector process $%
\dot{{\bf Y}}=(\dot{Y}_{1},\dot{Y}_{2},\dot{Y}_{3})$ is a commutative one
(self-nondemolition), 
\begin{equation}
\lbrack \dot{Y}_{i}(r),\dot{Y}_{k}(s)]=0  \label{eq:3}
\end{equation}%
and satisfies the nondemolition principle \cite{bib:ref11,bib:ref14} 
\begin{equation}
\lbrack \dot{Y}_{i}(t),\hat{Z}(s)]=0,\quad t\leq s,  \label{eq:4}
\end{equation}%
with respect to all future Heisenberg operators $\hat{Z}(t)=U(t)^{\ast }ZU(t)
$ of the particle. This means that the unitary evolution $U(t)$ can no
longer be the resolving operator for eq. (\ref{eq:2}) but must be defined
for an extended quantum system involving an apparatus with field coordinate
described by the white noise ${\bf e}$. A very nice model of such an
extended unitary evolution is based on the quantum stochastic Schr\"{o}%
dinger equation 
\begin{equation}
{\rm d}U+KU{\rm d}t=(d{\bf B}^{+}{\bf L}-{\bf L}^{+}{\rm d}{\bf B})U,
\label{eq:5}
\end{equation}%
$K={\bf L}^{+}{\bf L}/2+{\rm i}H/\hbar $, introduced by Hudson and
Parthasarathy in ref. \cite{bib:12}. Here $H$ is the Hamiltonian which in
the spinless case is 
\begin{equation}
H=[e{\bf A}(t,{\bf x})/c+{\rm i}\hbar {\bf \nabla }]^{2}/2m+e\Phi (t,{\bf x}%
),  \label{eq:6}
\end{equation}%
${\bf L}(t)$ and ${\bf B}(t)$ are the vector-operators (columns) of the
particle and of a Bose field respectively in the Schr\"{o}dinger picture, $%
{\bf L}^{+}=(L_{j}^{\ast })$ and ${\bf B}^{+}=(B_{j}^{\ast })$ are rows of
conjugate operators, ${\bf L}^{+}{\bf L}=\sum L_{j}^{\ast }L_{j}$, ${\bf B}%
^{+}{\bf L}=\sum B_{j}^{\ast }L_{j}$, ${\bf L}^{+}{\bf B}=\sum L_{j}^{\ast
}B_{j}$. The Bose field operators $B_{j}(t)$ are defined in Fock space as
annihilations by the canonical commutations relations 
\begin{eqnarray*}
\lbrack B_{i}(r),B_{k}(s)]\! &=&\!0, \\
\lbrack \dot{B}_{i}(r),\dot{B}_{k}^{\ast }(s)]\! &=&\!\delta _{ik}\delta
(r-s)
\end{eqnarray*}%
for the (generalized) derivations $\dot{B}_{j}(t)={\rm d}B_{j}/{\rm d}t$,
and ${\rm d}{\bf B}$ in (\ref{eq:5}) are the forward increments ${\rm d}{\bf %
B}(t)={\bf B}(t+{\rm d}t)-{\bf B}(t)$.

Let us respect the error in (\ref{eq:2}) as the operator-valued
vector-process 
\[
{\bf e}(t)=2\Re \dot{{\bf B}}(t)=\dot{{\bf B}}(t)+\dot{{\bf B}}(t)^{\ast },
\]%
having the correlations of the standard white noise, 
\[
\langle e_{i}(r)e_{k}(s)\rangle =\langle \dot{B}_{i}(r)\dot{B}_{k}^{\ast
}(s)\rangle =\delta _{ik}\delta (r-s),
\]%
with respect to the vacuum state of the Bose field. One can easily prove
that the nondemolition principle (\ref{eq:3}), (\ref{eq:4}) is fulfilled if $%
{\bf L}(t)=(2\lambda )^{1/2}{\bf X}$ where ${\bf X}$ is the coordinate
vector-operator of the particle given in the Schr\"{o}dinger representation
as multiplication by ${\bf x}=(x_{1},x_{2},x_{3})$. Indeed, in this case $%
{\bf Y}(t)=\int_{0}^{t}\dot{{\bf Y}}(r){\rm d}r$ is an output process ${\bf Y%
}=\hat{{\bf Q}}$ with respect to the evolution $U$ in the sense of refs. 
\cite{bib:ref14,bib:XP74,bib:ref18,bib:b5}: 
\[
\hat{Q}_{j}(t)=U(s)^{\ast }Q-j(t)U(s),\qquad s\leq t,\;\;j=1,2,3,
\]%
where 
\[
{\bf Q}(t)=\int\limits_{0}^{t}{\bf e}(r){\rm d}r=2\Re {\bf B}(t)
\]%
is the standard Wiener process in Fock space. Hence $\dot{Y}_{j}(t)$
commutes with $\dot{Y}_{k}(s)$ and $\hat{Z}(s)$ for all $s\geq t$ due to the
commutativity of $e_{j}(t)$ with $e_{k}(s)$ and $Z$. The corresponding
quantum Langevin equation 
\begin{equation}
{\rm d}\hat{Z}+(\hat{Z}\hat{K}+\hat{K}^{\ast }\hat{Z}-\hat{{\bf L}}^{+}Z{\bf 
\hat{L}}){\rm d}t={\rm d}{\bf B}^{+}[\hat{Z},{\bf \hat{L}}]+[{\bf \hat{B}}%
^{+},\hat{Z}]{\rm d}{\bf B},  \label{eq:7}
\end{equation}%
where $\hat{K}(t)=U(t)^{\ast }KU(t)$, $\hat{L}_{j}(t)=U(t)^{\ast }L_{j}U(t)$%
, gives the quantum stochastic Lorentz equation for $\hat{Z}=\hat{X}_{j}$, $%
j=1,2,3$, 
\[
\displaylines{
  m\hat{\ddot{\bm X}}=e(\hat{\bm E}+\hat{\dot{\bm X}}\times\hat{\bm H}/c)
  +\hbar(2\lambda)^{1/2}\Im\dot{\bm B}^*,\cr
  \hat{\bm E}=-\bm\nabla\Phi(\hat{\bm X}),\quad
  \hat{\bm H}=\bm\nabla\times\bm A(\hat{\bm X}), \cr
  (\hat{\dot{\bm X}}\times\hat{\bm H})_i=
  \epsilon_{ijk}(\hat{\dot X}_j\hat H_k+\hat H_k\hat{\dot X}_j)/2.}
\]

Note that the extra stochastic force 
\[
{\bf f}(t)=\hbar (2\lambda )^{1/2}\Im \dot{{\bf B}}^{\ast }=\frac{\hbar }{%
{\rm i}}\,(\lambda /2)^{1/2}(\dot{{\bf B}}^{\ast }-\dot{{\bf B}})
\]%
perturbs the Hamiltonian dynamics of the particle and to the observation (%
\ref{eq:2}) is another white noise of intensity $\hbar ^{2}\lambda /2$,
which does not commute with the error ${\bf e}$: 
\begin{equation}
\lbrack e_{j}(t),f_{k}(s)]=\frac{\hbar }{{\rm i}}\,(2\lambda )^{1/2}\delta
_{jk}\delta (t-s).  \label{eq:8}
\end{equation}%
Due to the openness of the observed particle as a quantum system in the Bose
reservoir, its prior dynamics is a mixing described by the irreversible
master equation 
\begin{equation}
\frac{{\rm d}}{{\rm d}t}\,\langle Z\rangle _{t}+\langle ZK+K^{\ast }Z-{\bf L}%
^{+}Z{\bf L}\rangle _{t}=0,  \label{eq:9}
\end{equation}%
obtained by averaging (\ref{eq:7}) with respect to the product $\varphi
_{0}=\psi \otimes |0\rangle $ of an initial wave function $\psi $ of the
particle and the vacuum state $|0\rangle $ of the Bose field. This means
that the prior expectations $\langle Z\rangle _{t}=\langle \varphi
_{t}|Z\varphi _{t}\rangle $ of the particle observables $Z$ for $\varphi
_{t}=U(t)\varphi _{0}$ cannot be described in terms of a wave function
involving only the particle, in spite of the fact that the initial state of
the particle is a pure one.

The posterior dynamics of the particle is described by posterior mean values 
$\hat{z}(t)=\langle Z\rangle ^{t}$, which are defined as conditional
expectations 
\begin{equation}
\langle Z\rangle ^{t}=\epsilon ^{t}(U(t)^{\ast }ZU(t))=\epsilon ^{t}(\hat{Z}%
(t))  \label{eq:10}
\end{equation}%
of $\hat{Z}=U^{\ast }ZU$ with respect to the observables ${\bf \hat{Q}}%
^{t}=\{{\bf Q}(r)|r\leq t\}$ up to the current time instant $t$ and the
initial state vector $\varphi _{0}$. As it was proved in ref. \cite{bib:XP74}%
, the nondemolition principle (\ref{eq:3}), (\ref{eq:4}) is necessary and
sufficient for the existence of the conditional expectation $\epsilon ^{t}:\ 
\hat{Z}\mapsto \epsilon ^{t}(\hat{Z})$ with respect to any initial $\varphi
_{0}$. For a fixed $\hat{Z}$, it is a non-anticipating $c$-valued function $%
\hat{z}:\ {\bf q}\mapsto z(t,{\bf q})=\langle Z\rangle ({\bf q}^{t})$ on the
observed trajectories ${\bf q}=\{{\bf q}(t)\}$ of the output process ${\bf Y}%
={\bf \hat{Q}}$; the conditional expectation $\epsilon ^{t}$ must satisfy
the positive projection conditions 
\[
\epsilon ^{t}(\hat{Z}^{\ast }\hat{Z})\geq 0,\quad \epsilon ^{t}\bigl(z(t,%
{\bf \hat{Q}})\bigr)({\bf q})=z(t,{\bf q}).
\]%
Hence for any particle operator $Z$ the posterior process $\hat{z}%
(t)=\langle Z\rangle ^{t}$ is a classical (commutative) stochastic one; its
averaging over all observable trajectories ${\bf q}$ coincides with the
prior mean value: $\langle \hat{z}(t)\rangle =\langle Z\rangle _{t}$. As a
linear map $Z\mapsto \langle Z\rangle ^{t}$ it is described by the quantum
filtering equation 
\begin{equation}
{\rm d}\langle Z\rangle ^{t}+\langle ZK+K^{\ast }Z-{\bf L}^{+}Z{\bf L}%
\rangle ^{t}{\rm d}t=\langle \tilde{Z}{\bf L}+{\bf L}^{+}\tilde{Z}\rangle
^{t}{\rm d}{\bf \hat{Q}},  \label{eq:11}
\end{equation}%
obtained in ref. \cite{bib:ref14} for the case considered 
\[
{\bf Y}(t)=\int\limits_{0}^{t}2\Re \lbrack {\bf L}(r){\rm d}r+{\rm d}{\bf B}%
]={\bf \hat{Q}}(t)
\]%
and ref. \cite{bib:XP74} for general output nondemolition processes with
respect to the initial vacuum Bose state. Here $\tilde{Z}(t)=Z-\hat{z}(t)I$
is the deviation of $Z$ in the Schr\"{o}dinger picture from the posterior
mean value $\hat{z}(t)=\langle Z\rangle ^{t}$ and 
\begin{equation}
{\bf \hat{Q}}(t)={\bf \hat{Q}}(t)-\int\limits_{0}^{t}[{\bf \hat{I}}(r)+{\bf 
\hat{I}}^{\ast }(r)]{\rm d}r  \label{eq:12}
\end{equation}%
is the observed innovating Wiener process, ${\bf \hat{I}}(t)({\bf q}%
)=\langle {\bf L}\rangle ({\bf q}^{t})$.

Let us prove that for any initial wave function $\psi ({\bf x})$ of the open
particle the posterior state (\ref{eq:9}) is pure and is given by 
\begin{equation}
\hat{z}(t)=\int \hat{\varphi}(t,{\bf x})^{\ast }Z\hat{\varphi}(t,{\bf x})%
{\rm d}{\bf x}=\hat{\varphi}(t)^{+}Z\hat{\varphi}(t),  \label{eq:13}
\end{equation}%
where the posterior wave function $\hat{\varphi}(t,{\bf x})({\bf q})=\varphi
({\bf q}^{t},{\bf x})$ satisfies the stochastic wave equation 
\begin{equation}
{\rm d}\hat{\varphi}+\tilde{K}\hat{\varphi}{\rm d}t=\tilde{L}\hat{\varphi}%
{\rm d}\tilde{{\bf Q}},\quad \hat{\varphi}(0,{\bf x})=\psi ({\bf x}).
\label{eq:14}
\end{equation}%
Indeed, if $\hat{\varphi}$ satisfies eq. (\ref{eq:14}) in the Ito form, then 
$\langle Z\rangle =\hat{\varphi}^{+}Z\hat{\varphi}$ satisfies the following
equation, 
\begin{equation}
{\rm d}\langle Z\rangle =\langle Z\tilde{K}+\tilde{K}^{\ast }\tilde{Z}-%
\tilde{{\bf L}}^{+}Z\tilde{{\bf L}}\rangle {\rm d}t=\langle Z\tilde{{\bf L}}+%
\tilde{{\bf L}}^{+}Z\rangle {\rm d}\tilde{{\bf Q}},  \label{eq:15}
\end{equation}%
obtained by using Ito's formula 
\begin{equation}
{\rm d}(\hat{\varphi}^{+}Z\hat{\varphi})={\rm d}\hat{\varphi}^{+}Z\hat{%
\varphi}+\hat{\varphi}^{+}Z{\rm d}\hat{\varphi}+{\rm d}\hat{\varphi}^{+}Z%
{\rm d}\hat{\varphi}  \label{eq:16}
\end{equation}%
and the Ito multiplication table ${\rm d}\tilde{Q}_{k}{\rm d}\tilde{Q}%
_{l}=\delta _{k}l{\rm d}t$. Comparing (\ref{eq:11}) and (\ref{eq:15}) and
taking into account the relation $\langle (Z-\hat{z}){\bf L}\rangle =\langle
Z({\bf L}-{\bf \hat{I}})\rangle $, one obtains $\tilde{K}={\bf \hat{L}}^{+}%
{\bf \hat{L}}/2+{\rm i}\tilde{H}/\hbar $ with 
\begin{equation}
{\bf \hat{L}}={\bf L}-\Re {\bf \hat{I}}+{\rm i}{\bf \hat{r}}/\hbar ,
\label{eq:17}
\end{equation}%
where ${\bf \hat{r}}(t)=(\hat{r}_{1},\hat{r}_{2},\hat{r}_{3})(t)$ and $\hat{s%
}(t)$ are arbitrary (inessential) real functions of ${\bf q}^{t}$ and 
\begin{equation}
\tilde{H}=H-\hbar \Re \hat{l}\Im {\bf L}-\hat{{\bf r}}\Re {\bf L}-\hat{s}
\label{eq:18}
\end{equation}%
is the Hamiltonian of the particle. Putting ${\bf \hat{r}}=0,\;\hat{s}=0$,
we obtain the following stochastic dissipative equation, 
\begin{equation}
{\rm d}\hat{\varphi}+({\bf \hat{L}}^{+}\tilde{{\bf L}}/2+{\rm i}\tilde{H}%
/\hbar )\hat{\varphi}{\rm d}t=\tilde{{\bf L}}\hat{\varphi}{\rm d}{\bf \hat{Q}%
},  \label{eq:19}
\end{equation}%
which is nonlinear because ${\bf \hat{L}}$ and $\tilde{H}$ depend on $\hat{%
\varphi}$ (\ref{eq:17}). Multiplying the posterior normalized wave function $%
\hat{\varphi}(t,{\bf x})$ by the stochastic amplitude $\hat{c}(t)$ which
satisfies the Ito equation 
\begin{equation}
{\rm d}\hat{c}+(\Re {\bf \hat{I}})^{2}\hat{c}{\rm d}t/2=(\Re {\bf \hat{I}})%
\hat{c}{\rm d}{\bf \hat{Q}},\quad \hat{c}(0)=1,  \label{eq:20}
\end{equation}%
using Ito's formula one can easily obtain 
\begin{eqnarray*}
{\rm d}(\hat{c}\hat{\varphi}) &=&\{\Re {\bf \hat{I}}{\rm d}{\bf \hat{Q}}%
-(\Re {\bf \hat{I}})^{2}{\rm d}t/2+{\bf \hat{L}}{\rm d}{\bf \hat{Q}}-\tilde{K%
}{\rm d}t+{\bf \hat{L}}\Re {\bf \hat{I}}{\rm d}t\}\hat{c}\hat{\varphi} \\
&=&\{(\Re {\bf \hat{I}}+{\bf \hat{L}}){\rm d}{\bf \hat{Q}}-[(\Re {\bf \hat{I}%
})^{2}/2+\tilde{K}+{\bf \hat{L}}\Re {\bf \hat{I}}]{\rm d}t\}\hat{c}\hat{%
\varphi}=({\bf L}{\rm d}{\bf \hat{Q}}-K{\rm d}t)\hat{c}\hat{\varphi},
\end{eqnarray*}%
where the relation ${\rm d}\hat{Q}_{l}{\rm d}\tilde{Q}_{k}=\delta _{lk}{\rm d%
}t$ for ${\rm d}{\bf \hat{Q}}={\rm d}{\bf \hat{Q}}-2\Re {\bf \hat{I}}{\rm d}t
$, ${\bf \hat{L}}={\bf L}-\Re {\bf \hat{I}}$, $\tilde{K}=K-{\bf L}\Re {\bf 
\hat{I}}+(\Re {\bf \hat{I}})^{2}/2$ was taken into account. Hence the
nonnormalized posterior wave function $\hat{\chi}(t,${${\bf x}$}$)=\hat{c}(t)%
\hat{\varphi}(t,${${\bf x}$}$)$ satisfies the linear stochastic equation 
\begin{equation}
{\rm d}\hat{\chi}+({\bf L}^{+}{\bf L}/2+{\rm i}H/\hbar )\hat{\chi}{\rm d}t=%
{\bf L}\hat{\chi}{\rm d}{\bf \hat{Q}}.  \label{eq:21}
\end{equation}%
The last equation can be transformed to a nonstochastic linear equation for $%
\hat{\psi}(t)=\exp [-{\bf L\hat{Q}}(t)]\hat{\chi}(t)$: 
\begin{equation}
{\rm i}\hbar \partial \hat{\psi}(t)/\partial t=H(\hat{Q}(t))\hat{\psi}%
(t),\quad \hat{\psi}(0)=\psi ,  \label{eq:22}
\end{equation}%
where 
\[
{\rm i}H({\bf Q}(t))=\hbar \exp [-{\bf L\hat{Q}}(t)](K+{\bf L}^{2}/2)\exp [%
{\bf LQ}(t)]
\]%
is a perturbed Hamiltonian $W(t)HW^{\ast }(t)$, $W(t)=\exp [-{\bf L\hat{Q}}%
(t)]$ ($W^{\ast }=W^{-1}$, if ${\bf L}^{\ast }=-{\bf L}$). Indeed, with the
help of Ito's formula, we obtain 
\begin{eqnarray*}
{\rm d}\hat{\psi} &=&{\rm e}^{-{\bf L\hat{Q}}}{\rm d}\hat{\chi}+{\rm d}{\rm e%
}^{-{\bf L\hat{Q}}}\hat{\chi}-{\bf L}^{2}{\rm e}^{-{\bf L\hat{Q}}}\hat{\chi}%
{\rm d}t \\
&=&{\rm e}^{-{\bf L\hat{Q}}}({\bf L}{\rm d}{\bf \hat{Q}}-K{\rm d}t)\hat{\chi}%
-({\bf L}{\rm d}{\bf \hat{Q}}-{\bf L}^{2}{\rm d}t/2){\rm e}^{-{\bf L\hat{Q}}}%
\hat{\chi}-{\bf L}^{2}\hat{\psi}{\rm d}t \\
&=&-{\rm e}^{-{\bf L\hat{Q}}}(K+{\bf L}^{2}/2){\rm e}^{{\bf L\hat{Q}}}\hat{%
\psi}{\rm d}t=\frac{1}{{\rm i}\hbar }H({\bf \hat{Q}}(t))\hat{\psi}{\rm d}t.
\end{eqnarray*}%
Eq. (\ref{eq:22}) for any observed trajectory ${\bf q}(t)$ can be viewed as
the Schr\"{o}dinger equation (\ref{eq:1}) with complex potentials ${\bf U}$
and $V$. In the case of position observation, ${\bf L}=(\lambda /2)^{1/2}%
{\bf x}$, these potentials have the form 
\begin{eqnarray}
{\bf U}(t,{\bf x}) &=&eA(t,{\bf x})/c+{\rm i}\hbar (\lambda /2)^{1/2}{\bf q}%
(t),  \nonumber \\
V(t,{\bf x}) &=&e\hat{\Phi}(t,{\bf x})-{\rm i}\hbar \lbrack (\lambda
/2)^{1/2}{\bf x}]^{2},  \label{eq:23}
\end{eqnarray}%
and we get 
\begin{eqnarray*}
H({\bf q}) &=&\exp [-(\lambda /2)^{1/2}{\bf xq}](H-\frac{1}{2}{\rm i}\hbar
\lambda {\bf x}^{2})\exp [{\bf xq}(\lambda /2)^{1/2}= \\
&&\frac{1}{2m}\bigl\{(e/c)A+[\nabla -(\lambda /2)^{1/2}{\bf q}]^{2}\bigr\}%
+e\Phi +\frac{\hbar \lambda }{2{\rm i}}{\bf x}^{2}.
\end{eqnarray*}%
The solution $\hat{\psi}(t,{\bf x})({\bf q})=\psi ({\bf q}^{t},{\bf x})$ of
the wave equation (\ref{eq:22}) for an observed trajectory ${\bf q}=\{{\bf q}%
(t)\}$ defines the posterior normalized wave function 
\begin{equation}
\varphi ({\bf q}^{t},{\bf x})=\exp [(\lambda /2)^{1/2}{\bf xq}(t)-\ln c({\bf %
q}^{t})]\psi ({\bf q}^{t},{\bf x}),  \label{eq:24}
\end{equation}%
where 
\[
\ln \hat{c}(t)=\int_{0}^{t}(\lambda /2)^{1/2}\hat{{\bf x}}{\rm d}\hat{{\bf Q}%
}-\lambda \hat{{\bf x}}^{2}{\rm d}t/2
\]%
can be obtained from the normalization condition 
\begin{equation}
c({\bf q}^{t})^{2}=\int \exp [(2\lambda )^{1/2}{\bf xq}(t)]|\psi ({\bf q}%
^{t},{\bf x})|^{2}{\rm d}{\bf x}.  \label{eq:25}
\end{equation}

Note that the indirect nondemolition measurement considered is complete in
the sense that the posterior state of the particle is pure if the initial
state is pure. Hence the posterior dynamics of such indirectly completely
observed particle is pure contrary to the prior dynamics which is always
mixed (for $\lambda\neq 0$) even for vacuum quantum noise. In the case of
noncomplete measurement, if for instance, an extra bath is added, or if the
noise has a nonzero temperature, as is supposed in ref. \cite{bib:ref11},
this fact is no more true.

The author is grateful to Professor R. S. Ingarden and Dr. P. Staszewski for
their hospitality and useful discussions during his stay at the Institute of
Physics, N. Copernicus University, Toru\'{n}, where this work was completed.

\bibliographystyle{plain}
%

\end{document}